# Spatially Dependent in-Gap States Induced by Andreev Tunneling through a Single Electronic State

Ruixia Zhong[1,*], Zhongzheng Yang[1,*], Qi Wang[1,*], Fanbang Zheng[1], Wenhui Li[1], Juefei Wu[1], Chenhaoping Wen[1], Xi Chen[2], Yanpeng Qi[1,3,4,†], Shichao Yan[1,3,†]

[1] *School of Physical Science and Technology, ShanghaiTech University, Shanghai 201210, China*

[2] *State Key Laboratory of Low-Dimensional Quantum Physics, Department of Physics, Tsinghua University, Beijing 10084, China*

[3] *ShanghaiTech Laboratory for Topological Physics, ShanghaiTech University, Shanghai 201210, China*

[4] *Shanghai Key Laboratory of High-resolution Electron Microscopy, ShanghaiTech University, Shanghai 201210, China*

[*] *These authors contributed equally to this work*

[†] *Email*: yanshch@shanghaitech.edu.cn; qiyp@shanghaitech.edu.cn

**ABSTRACT**

By using low-temperature scanning tunneling microscopy and spectroscopy (STM/STS), we observe in-gap states induced by Andreev tunneling through a single impurity state in a low carrier density superconductor (NaAlSi). The energy-symmetric in-gap states appear when the impurity state is located within the superconducting gap. In-gap states can cross the Fermi level, and they show X-shaped spatial variation. We interpret the in-gap states as a consequence of the Andreev tunneling through the impurity state, which involves the formation or breakup of a Cooper pair. Due to the low carrier density in NaAlSi, the in-gap state is tunable by controlling the STM tip-sample distance. Under strong external magnetic fields, the impurity state shows Zeeman splitting when it is located near the Fermi level. Our findings not only demonstrate the Andreev tunneling involving single electronic state, but also provide new insights for understanding the spatially-dependent in-gap states in low carrier density superconductors.

In-gap states may arise in superconductors when the superconducting order parameter varies in space[1]. For instance, magnetic impurities in s-wave superconductors can lead to Yu-Shiba-Rusinov (YSR) state[2-8], and weak links between two superconductors can generate Andreev states[9]. Due to the protection of the superconducting gap, in-gap states also provide a promising platform for applications in quantum computing[10-12]. Tunneling experiments can provide a wealth of information on in-gap states, such as the energy position, orbital structure, and quasiparticle relaxation through the in-gap states[6, 13-14]. Andreev tunneling, which involves the creation or annihilation of a Cooper pair, is a key process when tunneling into or out of a superconductor[15-16]. Tunneling through YSR state can be considered as carried by the Andreev process[14, 17]. However, observing the Andreev tunneling through a pure electronic state within the superconducting gap is challenging. This is because superconductivity often emerges in high carrier density systems where the single electronic state could be strongly screened. The screening effect is reduced in low carrier density superconductors[18-19], which may make it possible to observe the Andreev tunneling involving single electronic state.

NaAlSi is an intrinsically self-doped low carrier density superconductor with the superconducting transition temperature ($T_c$) at ~7 K[20-25], and its electron carrier density is estimated to be about $6.4 \times 10^{20}$ cm$^{-3}$ at 10 K[25]. More interestingly, topological nodal surfaces and nodal ring states have also been recently observed in NaAlSi[26]. Here we report a low-temperature scanning tunneling microscopy and spectroscopy (STM/STS) study on NaAlSi. We detect the energy-symmetric in-gap states induced by Andreev tunneling through a single impurity state. We find that the in-gap states in NaAlSi show strong spatial variation, and investigate the impurity state behavior under external magnetic fields. We also discover that due to the low carrier density in NaAlSi, this kind of in-gap state is highly tunable.

As shown in Figure 1a, NaAlSi has the same structure as the "111" type ion-pnictide superconductors[20-21]. Cleaving NaAlSi single crystals results in a Na-terminated surface with a square lattice. Figure 1c,d are the typical constant-current STM topographies taken on the surface of NaAlSi. There are two main types of atomic defects (Figure 1c,d and Figure S1). One defect appears as a dark hole, and the other defect is cross-shaped (as shown by the arrows in Figure 1c,d). The dark hole may indicate that this defect is Na vacancy. In the zoom-in STM topography with cross-shaped defects (Figure 1d), we can still observe the square lattice of the Na atoms, which shows the cross-shaped defects are likely from the underneath AlSi layer.

We first characterize superconductivity in NaAlSi by measuring the temperature and magnetic field dependent differential conductance (d$I$/d$V$) spectra. As shown in Figure 1e, there is full-gap superconducting spectrum at ~130 mK, and it is gradually suppressed as the temperature increases. In order to quantify the size of the superconducting gap, we fit the temperature-dependent d$I$/d$V$ spectra with an isotropic s-wave gap function (see Figure S2), and the temperature dependence of the superconducting gap ($\Delta$) shows a BCS-gap behavior (Figure 1f)[27-28]. The extracted $T_c$ for NaAlSi is about ~7 K, which is consistent with the $T_c$ obtained in the electrical transport measurement (Figure 1b). The low-temperature superconducting gap is about 1 meV, which yields a ratio of $2\Delta/k_B T_c \approx 3.32$. In addition to the temperature-dependent d$I$/d$V$ spectra, we also carry out d$I$/d$V$ measurements with the magnetic fields perpendicular to the NaAlSi surface. As the strength of the magnetic fields increases, the superconducting gap feature gradually disappears (Figure 1g). Fitting the magnetic-field-dependent superconducting gap with $\Delta(H) = \Delta(0)[1 - (H/H_{c2})^2]^{1/2}$ results in an upper critical field about $0.8 \pm 0.1$ T (Figure 1h).

Although we can measure the full-gap superconducting spectrum on the NaAlSi surface (Figure 1e), there are ring-shaped features in the d$I$/d$V$ maps taken with the bias voltages within the superconducting gap (Figure 2b,c and Figure S3). This indicates that there are in-gap states within the superconducting gap near the ring-shaped areas. In this manuscript, we use "ring" to label the ring-shaped feature in the d$I$/d$V$ maps for simplicity. We find that the individual atomic defects as shown in the STM topography are not directly related with in-gap states (see Figure S4). In comparison with the concentration of the defects on the surface of NaAlSi, the rings shown in the d$I$/d$V$ maps are sparse, and there are roughly 15 rings in a 120 by 120 nm$^2$ area (see Figure S3). At the center of the rings in the d$I$/d$V$ maps (Figure 2a,b and Figure S5), there are often impurity features, which indicates that the impurity states are likely induced by the subsurface impurities in NaAlSi. We also note that no clear vortex-core state is observed in NaAlSi (see Figure S6).

Figure 2d shows the d$I$/d$V$ linecut profile taken across the ring in Figure 2b, and Figure 2f shows the typical d$I$/d$V$ spectra marked by the colored arrows in Figure 2d. We can see that there are three kinds of areas: (1) Inside the ring, besides the superconducting gap, there is an electronic state that is located above the superconducting coherence peak (spectrum 1 in Figure 2f). We name this electronic state as "impurity state". When moving toward the ring, this impurity state gradually shifts toward the Fermi level (spectrum 2 in Figure 2f); (2) Near the ring, there are sharp energy-symmetric resonances inside the superconducting gap that extend over ~5 nm in real space (spectra 3-5 in Figure 2f). Importantly, in this area, the energies of the in-gap states show strong spatial variation. The sub-gap states shift toward the Fermi level, then they cross at the Fermi level and split again. This results in the X-shaped in-gap states (Figure 2d); (3) Outside the ring, the full-gap superconducting spectrum recovers (spectrum 6 in Figure 2f).

To reveal the influence of superconductivity on the in-gap states, we apply external magnetic fields above the critical magnetic field to suppress superconductivity in NaAlSi. Figure 2e is the d$I$/d$V$ linecut profile taken with 1.5 T magnetic field along the same arrow as for Figure 2d (see Figure S7). Figure 2g shows the d$I$/d$V$ spectra in Figure 2e taken at the same positions as the corresponding spectra in Figure 2f. We can see that as long as superconductivity is suppressed by the external magnetic fields, the sharp energy-symmetric in-gap states disappear, and there is a broader impurity state near the Fermi level (spectra 3-5 in Figure 2g). This indicates that the in-gap states emerge where the impurity state is located within the superconducting gap. The width of the in-gap states is significantly reduced compared to that of the impurity state (Figure 2d). This is due to the negligible scattering into the gapped bulk states, the lifetime of the impurity state located within the superconducting gap is greatly enhanced[29-30].

For different rings shown in the d$I$/d$V$ maps in Figure 2b and c, the relative energy positions between the impurity state and the Fermi level could be different. Figure 2h and i are the d$I$/d$V$ linecut profiles across another ring shown in Figure 2c with 0 T and 1.5 T magnetic fields, respectively. We can see that even inside the ring, the electronic state is still slightly below the Fermi level (Figure 2i). The energy-symmetric in-gap states appear around the center of the ring where the impurity state is located within the superconducting gap (Figure 2h). The different energy positions of the impurity state could be due to the different locations of the impurities in NaAlSi. We note that the impurity state in NaAlSi can be located below and above the Fermi level (Figure 2e,i). This is different as the Kondo resonance peak which is pinned at the Fermi level[31-32]. In the d$I$/d$V$ spectrum taken with larger energy range, in addition to the impurity state located around the Fermi energy, we can also observe the other discrete energy states which could be the higher-energy impurity states (see Figure S8). The energy level spacing for the discrete energy states is about 15 meV.

We next discuss the mechanism for observing the energy-symmetric in-gap states in NaAlSi. For s-wave superconductors, in-gap states are often attributed as YSR states which are due to the exchange scattering potential of magnetic impurities, and the YSR states do not require a pure electronic state[2-8]. The appearance of YSR states often results in a suppression in the intensity of the coherence peak[33-34]. In our case, the in-gap states and their energy positions are directly related with the impurity state (Figure 2e,i), and the intensities of the superconducting coherence peaks have no clear difference in the d$I$/d$V$ spectra with and without in-gap states (Figure 2d,f). These features are different from the conventional YSR states. We divide the in-gap states in NaAlSi into two scenarios: one scenario is that the impurity state ($V_p$) is slightly above the Fermi level ($0 < V_p < \Delta$, Figure 3a,b), and the other is the impurity state is slightly below the Fermi level ($-\Delta < V_p < 0$, Figure 3c,d). For the first scenario, when we apply positive sample bias voltage ($V_s > 0$), electrons tunnel from the STM tip to NaAlSi. The electrons can tunnel into the impurity state when $V_s$ equals $V_p$ (Figure 3a), and we could observe a peak feature in the d$I$/d$V$ spectrum. With negative sample bias voltage ($V_s < 0$), electrons tunnel from NaAlSi to the STM tip. When the absolute value of $V_s$ equals $V_p$, the Cooper pair breakup occurs in NaAlSi, one electron tunnels inelastically into the STM tip and excites the other electron into the impurity state (Figure 3b)[35]. Due to this Andreev tunneling process with the breakup of a Cooper pair, the impurity state located above the Fermi level can also contribute to the tunneling current even with negative bias voltage (Figure 3b).

For the second scenario (Figure 3c,d), the impurity state is located slightly below the Fermi level ($-\Delta < V_p < 0$), and it is singly occupied. When we apply negative sample bias voltage ($V_s < 0$), electrons tunnel from NaAlSi to STM tip. The electrons in the electronic state tunnel into STM tip when $V_s$ equals $V_p$ (Figure 3d), and we measure a peak feature at $V_s$ in the d$I$/d$V$ spectrum. With positive bias voltage ($V_s > 0$), the electrons tunnel from STM tip to NaAlSi. When $V_s$ equals the absolute value of $V_p$, one electron from STM tip tunnels inelastically into NaAlSi and excites the electron in the impurity state to form a Cooper pair in NaAlSi (Figure 3c)[35]. With this Andreev tunneling process involving the formation of a Cooper pair, we can also observe a peak feature in the positive bias voltage side of the d$I$/d$V$ spectrum.

Having understood the origin of the particle-hole symmetric in-gap states in NaAlSi, the next question is why the in-gap states exhibit spatial variation? First, our d$I$/d$V$ spectra data clearly indicate that the in-gap states are induced by the impurity state, and the energy positions of the sub-gap states are also determined by the relative energy position between the impurity state and the Fermi level. Second, due to the different materials, NaAlSi could have significantly different work function as the STM tip, which induces a voltage drop between NaAlSi and STM tip[36]. Third, because of the low carrier density in NaAlSi, there would be non-zero screening length that results in the penetration of the electric field from STM tip into NaAlSi. Based on the above three factors, we propose that the STM tip can behave as a gate electrode that tunes the energy levels of the impurity[36-40]. When moving the STM tip laterally toward the impurity location, the local electric field effect from the STM tip gets stronger. It reaches the maximum when the STM tip is right above the impurity. This can nicely explain the spatial dependence of the impurity state shown in the d$I$/d$V$ linecut profiles in Figure 2.

This gating effect of the STM tip can also be confirmed by tuning the impurity state with tip-sample distance. We control the tip-sample distance by using the STM-based constant-current feedback, and increase the tunneling current setpoint while keeping the bias voltage constant. Figure 4a shows three d$I$/d$V$ linecut profiles measured with different tip-sample distances along the arrow in the inset of Figure 4a (see Figure S9 for the typical d$I$/d$V$ spectra), and the tip-sample distance is measured relative to the setpoint: $V_s = -10$ mV, $I = 150$ pA. When the STM tip is brought closer to the NaAlSi surface, the local electric field becomes stronger and the impurity electronic state is shifted to higher energy. In order to see the influence of the tip-sample distance to the in-gap states, we measure the d$I$/d$V$ linecut profiles taken near the ring and with smaller energy scale (Figure 4b and S10). The tip-sample distance is measured relative to the setpoint: $V_s = -3$ mV, $I = 150$ pA. In this case, as reducing the tip-sample distance, the locations showing the X-shaped in-gap states and the crossing point at zero bias voltage appear at larger distance from the center of the ring (Figure 4b).

We also investigate the response of the impurity state to strong external magnetic fields that are significantly larger than the critical magnetic field in NaAlSi. Figure 4d is a d$I$/d$V$ map that shows another ring feature. Figure 4e-g are the evolution of the d$I$/d$V$ spectra under external magnetic fields taken at the locations marked in Figure 4d. Similar as the other rings shown in Figure 2, when moving toward the center of the ring, the energy position of the impurity state shifts away from the Fermi level (Figure 4e-g). When the impurity electronic state is located at the Fermi level, there is clear energy splitting under strong external magnetic field (Figure 4e). By performing a linear fit to the magnetic-field-induced energy splitting and assuming it is Zeeman-type, the extracted Landé $g$-factor for this impurity state is about $2.87 \pm 0.04$ (see Figure S11). When the impurity state is shifted away from the Fermi level, the magnetic-field-induced energy splitting effect is reduced (Figure 4f,g and S12). This could be because the lifetime of the impurity state at higher energy is shorter and its width becomes larger, which make the energy splitting less clear.

The X-shaped in-gap states have also been observed on the intrinsic impurities of FeTe$_{0.55}$Se$_{0.45}$ and in the Pb/Co/Si(111) sandwiched structures[36, 41]. In FeTe$_{0.55}$Se$_{0.45}$, they have been interpreted as YSR states induced by magnetic impurities[36], and they are suggested to be topological in the Pb/Co/Si(111) system[41]. However, these two works focus on the energy scales within the superconducting gap, and no clear impurity state is reported in these two systems. Very recently, it has been shown that when the quantum level of a surface state confined in a quantum corral is located close to the Fermi level, in-gap states appear within the superconducting gap, and the proximity effect would open a gap in the confined surface state[30]. According to the Hamiltonian used in Refs.[30, 42], the X-shaped in-gap states in NaAlSi indicate that the coupling between impurity state and superconducting electrons in NaAlSi is very weak.

In conclusion, by using high-resolution STM/STS, we observe in-gap states with strong spatial variation in NaAlSi. The energy-symmetric in-gap resonances are induced by Andreev tunneling through the localized impurity state. Due to the low carrier density in NaAlSi, the in-gap states can be tuned by the local gating effect from the STM tip, which also explains their spatial dependence. Our results demonstrate that the localized electronic state coupling weakly with the superconducting bath can induce in-gap states by Andreev tunneling process. The mechanism for the spatially-dependent in-gap states revealed here advances the understanding of the interplay between localized impurity state and superconductivity with low carrier density.

## MATERIALS AND METHODS

**Synthesis of NaAlSi:** NaAlSi single crystals were grown by Ga flux method. The mixture of elements loaded in an alumina crucible with the molar ratio of Na : Al : Si : Ga = 3 : 1 : 1 : 0.5 was sealed in a tantalum crucible and then put in a sealed quartz ampoule. The mixture was heated up to 1173 K for 12 hours, and then cooled down to 1073 K for 1 hour before dropping to 823 K at the rate of 2.5 K/hour. Finally, the quartz ampoule was cooled down to room temperature. The extra Na was removed by utilizing alcohol and the high-quality single crystals of NaAlSi were obtained.

**STM/STS Measurements:** STM/STS experiments were conducted with a Unisoku low-temperature and high-magnetic-field STM (USM1600). The tungsten tips were used in the STM measurements, and they were flashed by electron-beam bombardment for several minutes before use. NaAlSi samples were cleaved at 77 K and in an ultrahigh-vacuum chamber. After cleaving, the samples are immediately transferred into the low-temperature STM head for measurements. STS was measured using a standard lock-in detection technique with a frequency of 914 Hz.

## ASSOCIATED CONTENTS

**Supporting Information:** STM topographies taken on NaAlSi with various bias voltages; Fitting superconducting gap in NaAlSi; Ring features in the d$I$/d$V$ maps taken on a large area; In-gap states and the individual atomic defects in the STM topography; More d$I$/d$V$ maps with ring features; Superconducting vortices in NaAlSi; Comparison between d$I$/d$V$ maps without and with external magnetic fields; d$I$/d$V$ linecut profile with a large energy range; Typical d$I$/d$V$ spectra in Figure 4a; Typical d$I$/d$V$ spectra in Figure 4b; Extracting Landé $g$-factor from magnetic-field-dependent d$I$/d$V$ spectra; Linecut d$I$/d$V$ profiles under strong external magnetic fields

## AUTHOR INFORMATION

**Corresponding authors**
Correspondence and requests for materials should be addressed to S.Y. or Y.Q.
**Author contributions**
S.Y. conceived the experiments. R.Z., Z.Y., F.Z. and S.Y. carried out the STM experiments and experimental data analysis. Q.W. and Y.Q. were responsible for sample growth. X.C. developed the Andreev tunneling model. R.Z., W. Li, and S.Y. wrote the manuscript with input from all authors.
**Notes**
The authors declare no competing financial interest.

## ACKNOWLEDGEMENTS

S.Y. acknowledges the financial support from the National Key R&D program of China (2020YFA0309602, 2022YFA1402703) and the start-up funding from ShanghaiTech University. Y.Q. acknowledges the financial support from the National Key R&D program of China (2023YFA1607400, 2018YFA0704300) and the National Natural Science Foundation of China (52272265). W.L. acknowledges the financial support from the Postdoctoral Fellowship Program of CPSF (Grant No. GZC20231670).

**Figure 1**

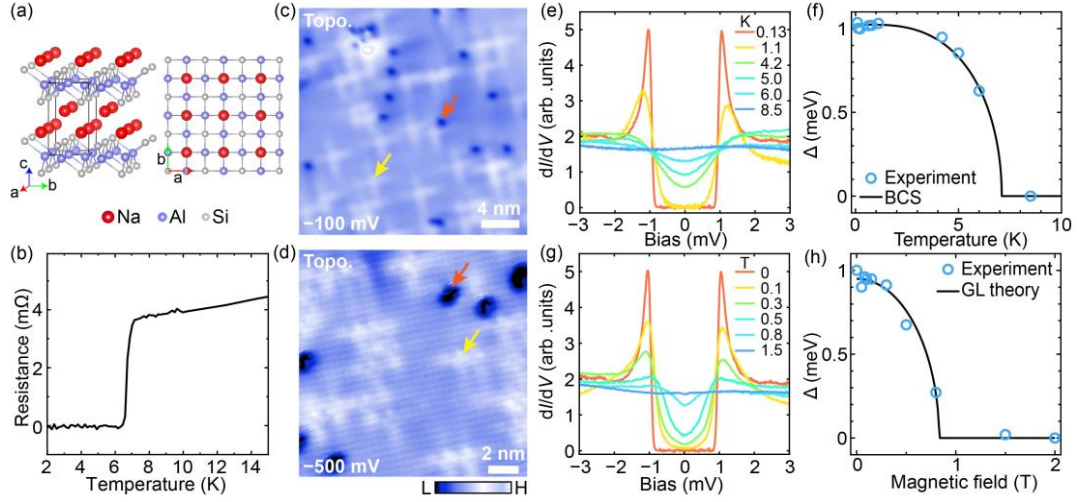

**Figure 1.** (a) Schematic showing the crystal structure of NaAlSi. (b) Temperature-dependent electrical resistance of NaAlSi, which shows a sharp superconducting transition at ~7 K. (c) Constant-current STM topography taken on the NaAlSi surface ($V_s = -100$ mV, $I = 100$ pA). (d) High-resolution STM topography taken on the NaAlSi surface ($V_s = -500$ mV, $I = 100$ pA). The orange and yellow arrows in (c) and (d) mark the two typical atomic impurities on the surface. (e, f) Temperature dependences of the d$I$/d$V$ spectra (e) and the size of the superconducting gap (f). The solid line in (f) is a BCS fitting for the superconducting gap. (g, h) Magnetic-field dependences of the d$I$/d$V$ spectra (g) and the size of the superconducting gap (h). The solid line in (h) is a calculated curve by using the Ginzburg-Landau (GL) theory. Error bars in (f) and (h) are comparable to the symbol size.

**Figure 2**

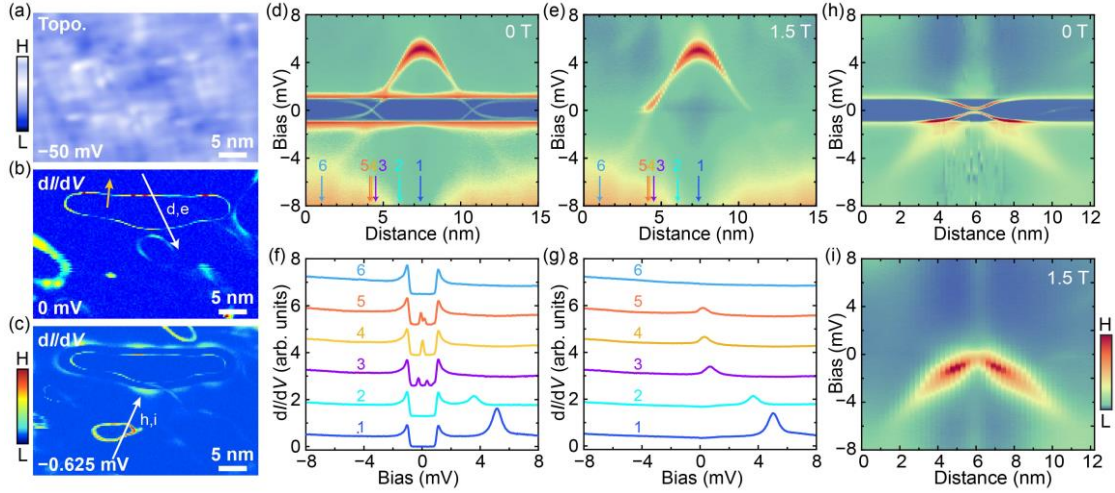

**Figure 2.** (a) Constant-current STM topography taken on the NaAlSi surface ($V_s = -50$ mV, $I = 100$ pA). (b, c) d$I$/d$V$ maps on the area shown in (a) with 0 mV and −0.625 mV bias voltages, respectively. (d, e) d$I$/d$V$ linecut profiles taken along the white arrow in (b) with 0 T (d) and 1.5 T (e) magnetic fields. Setup condition: $V_s = -10$ mV, $I = 300$ pA. (f, g) The typical d$I$/d$V$ spectra marked by the colored arrows in (d) and (e), respectively. (h, i) d$I$/d$V$ linecut profiles taken along the white arrow in (c) with 0 T (h) and 1.5 T (i) magnetic fields. Setup condition: $V_s = -10$ mV, $I = 300$ pA. The data shown in this figure are taken at 130 mK.

**Figure 3**

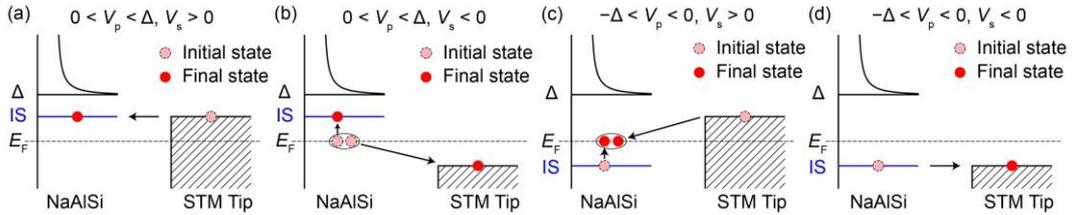

**Figure 3.** (a, b) The electron tunneling model when the impurity state (IS) in NaAlSi is located slightly above the Fermi level ($0 < V_p < \Delta$). $V_p$ denotes the energy position of the impurity state. (c, d) The electron tunneling model when the impurity state in NaAlSi is located slightly below the Fermi level ($-\Delta < V_p < 0$). For (a, c), a positive bias voltage ($V_s > 0$) is applied to NaAlSi, the electron tunnels from STM tip to NaAlSi. For (b, d), a negative bias voltage ($V_s < 0$) is applied to NaAlSi, the electron tunnels from NaAlSi to STM tip.

**Figure 4**

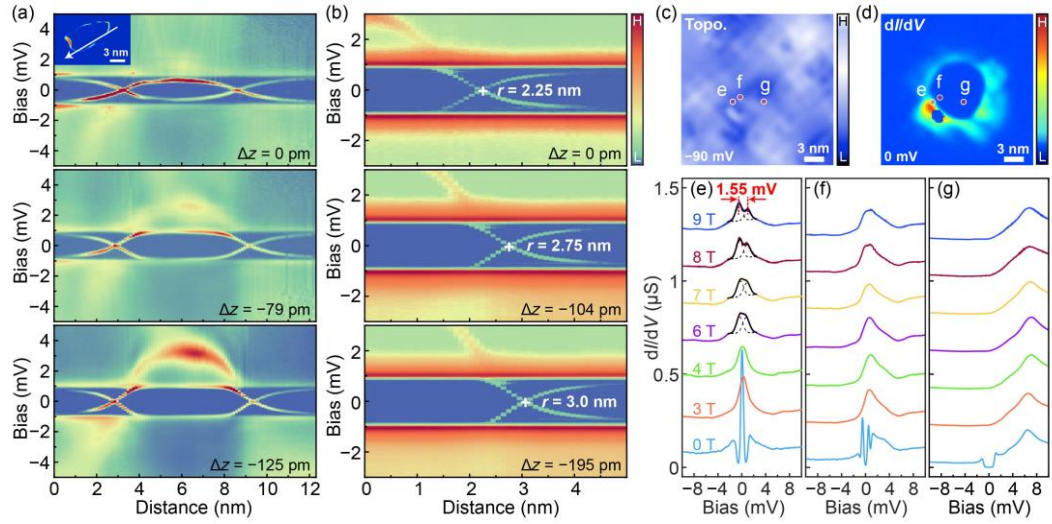

**Figure 4.** (a) d$I$/d$V$ linecut profiles taken along the arrow shown in the inset of (a) with different tip-sample distances ($\Delta z$) at 30 mK. The tip-sample distance is measured relative to the setpoint $V_s$ = −10 mV, $I$ = 150 pA. (b) d$I$/d$V$ linecut profiles taken along the orange arrow in Figure 2b with different tip-sample distances ($\Delta z$) at 130 mK. The tip-sample distance is measured relative to the setpoint $V_s$ = −3 mV, $I$ = 150 pA. The negative value for $\Delta z$ means the STM tip moves closer to the sample. "$r$" denotes the distance from the cross point at zero bias voltage to the starting point for the d$I$/d$V$ linecut. (c, d) Constant-current STM topography (c) and d$I$/d$V$ map taken with 0 mV bias voltage (d). (e-g) Magnetic-field dependent d$I$/d$V$ spectra on the positions marked in (c, d). Setup condition: $V_s$ = 10 mV, $I$ = 1.2 nA. The black curves are the fitting results with two Gaussian profiles. The d$I$/d$V$ spectra in (e-g) are vertically offset for clarity and taken at ~800 mK.